\documentclass[conference]{IEEEtran}
\IEEEoverridecommandlockouts

\usepackage{cite}
\usepackage{amsmath,amssymb,amsfonts}
\usepackage{algorithmic}
\usepackage{graphicx}
\usepackage{textcomp}
\usepackage{hyperref}
\usepackage{xcolor}
\usepackage{listings}
\usepackage{supertabular,booktabs}
\usepackage{enumitem}
\usepackage{todonotes} 
\usepackage{subcaption}
\usepackage{comment}
\def\BibTeX{{\rm B\kern-.05em{\sc i\kern-.025em b}\kern-.08em
    T\kern-.1667em\lower.7ex\hbox{E}\kern-.125emX}}
\begin{document}

\title{Dynamic Provisioning of REST APIs for Model Management}

\author{
    \IEEEauthorblockN{
        Adiel Tuyishime\IEEEauthorrefmark{1}, 
        Francesco Basciani\IEEEauthorrefmark{1}, 
        Javier Luis Cánovas Izquierdo\IEEEauthorrefmark{2} 
        and Ludovico Iovino\IEEEauthorrefmark{1}
    }
    \IEEEauthorblockA{\IEEEauthorrefmark{1}Gran Sasso Science Institute\\L'Aquila, Italy\\\{adiel.tuyishime,francesco.basciani,ludovico.iovino\}@gssi.it}
    \IEEEauthorblockA{\IEEEauthorrefmark{2}IN3 -- UOC\\Barcelona, Spain\\jcanovasi@uoc.edu}
}

\maketitle

\begin{abstract}
Model-Driven Engineering (MDE) is a software engineering methodology focusing on models as primary artifacts.
In the last years, the emergence of Web technologies has led to the development of Web-based modeling tools and model-based approaches for the Web that offer a web-based environment to create and edit models or model-based low-code solutions.
A common requirement when developing Web-based modeling tools is to provide a fast and efficient way for model management, and this is particularly a hot topic in model-based system engineering.
However, the number of approaches offering RESTful services for model management is still limited. 
Among the alternatives for developing distributed services, there is a growing interest in the use of RESTful services. 
In this paper, we present an approach to provide RESTful services for model management that can be used to interact with any kind of model, and can be used to build a modeling platform providing modeling-as-a-service.
The approach follows the REST principles to provide a stateless and scalable service.
\end{abstract}


\section{Introduction}
\label{sec:introduction}

Web services offer a suitable solution to provide efficient access to data storage. 
Among the alternatives for developing distributed services (e.g., SOAP, or WSDL), there is a growing interest in using RESTful services.
REST proposes stateless distributed services and relies on simple URIs and HTTP verbs to make Web services available to clients.

Model-driven Engineering (MDE) is a software engineering methodology focusing on models as primary artifacts.
The emergence of Web technologies has led to the development of Web-based modeling tools, such as Sirius Web\footnote{\url{ https://eclipse.dev/sirius/}} or Theia\footnote{\url{https://theia-ide.org/}}, which offer a web-based environment to create and edit models; and model-based approaches for the Web, such as WebRatio\footnote{https://www.webratio.com}, which provide a model-based low-code solution.

Web-based modeling tools require a way to manage the models created by the users.
Nowadays, the number of approaches offering RESTful services for model management is still limited, being EMF-REST~\cite{DBLP:conf/sac/Ed-DouibiIGTC16} one of them.
However, EMF-REST requires the provision of a model beforehand to provide a REST API.
This process may become a limitation in distributed modeling platforms, where the models are not known in advance and the API must be provided on the fly.

In this paper, we present an approach to dynamic provisioning RESTful services for model management, where a REST API is provided on-the-fly for any kind of model.
These services can interact with any kind of model and can be used to build a modeling platform providing modeling-as-a-service.
Our approach is currently a work in progress, but a first prototype has been implemented as a middleware, which can be used to expose RESTful services for model management.

The rest of the paper is organized as follows.
Section~\ref{sec:approach} presents our approach.
Section~\ref{sec:related} discusses the related work, and Section~\ref{sec:conclusion} concludes the paper and outlines future work.

\section{Our Approach}
\label{sec:approach}

This section presents our approach for provisioning a REST API middleware for a distributed modeling platform. 
We first describe the architecture of our approach, then the automated provisioning process to create REST API middleware, and finally, the supported operations by the middleware.

\subsection{Architecture}
\begin{figure*}[t]
    \centering
    \includegraphics[width=0.85\textwidth]{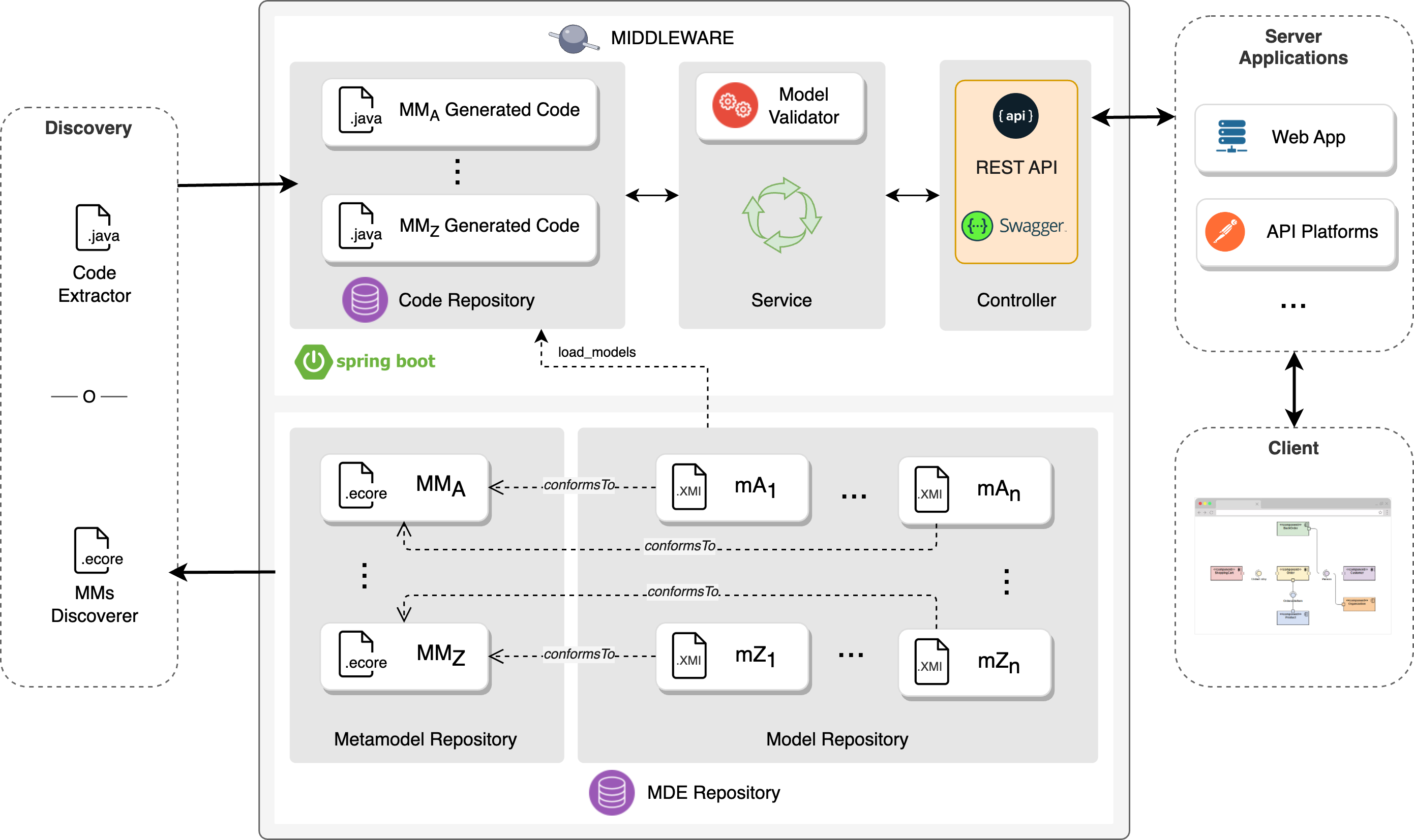}
    \caption{Architecture of our approach}
    \label{fig:architecture}
\end{figure*}

Figure~\ref{fig:architecture} shows the architecture of our approach. 
As can be seen, we identify an \emph{MDE repository}, which stores model-based artifacts; and a \emph{Middleware}, which offers a REST API to access the models stored within the MDE Repository.  

The MDE repository stores the models and metamodels of the domains of interest.
A metamodel can be defined as a domain description in terms of metaclasses, properties, and references; while models are specific instances of the metamodels, and comply with the structure and constraints defined in the metamodels.

The middleware follows a three-layer layered architecture, inspired by the Spring Boot framework\footnote{\url{ https://spring.io/projects/spring-boot}}, including code repository, service, and controller layers.
The code repository stores and manipulates data, thus encapsulating the fundamental CRUD operations for model management. Model operations are executed in stateless mode, thus ensuring that operations are self-contained and independent. 
The service layer bridges the code repository and the controller, thus digesting model data to be provided to the service.
This layer also includes a \emph{Model Validator}, which evaluates the correctness of the models (i.e., structural and domain constraints).
Finally, the controller layer provides the main interface for external applications, managing incoming requests and formulating the responses.
      
An example of an operational flow is described as follows.
The process begins at the controller layer, which handles incoming requests. 
The service layer processes the requests, often requiring data interactions with the code repository layer. 
Upon completion of the processing, the service layer relays the outcomes to the controller, which builds a response.

\subsection{Provisioning of REST APIs} 
The provisioning of REST APIs leverages on model reflection to dynamically provide a REST API tailored to the metamodels and models stored in the repository. 
The provisioning process is launched whenever any update is introduced into the MDE Repository. 
Upon detecting a metamodel update, the process performs two steps: (1) update of the generator model (\textit{GenModel}) and (2) generation of the corresponding Java code.
A \textit{GenModel} is a configuration model used by EMF\cite{steinberg2008emf} to generate Java code from metamodels that will be hosted in the code repository. 
Once the \textit{GenModel} is updated, the system regenerates the corresponding Java code, thus ensuring the new (or updated) metamodel is fully integrated in our middleware without the need to redeploy and restart the application.

\subsection{Supported operations by the middleware}
Model traversals are defined using generic REST API operations to enable this dynamic provisioning (e.g., the operation $/MM_A/ClassName/all$ get all the instances of the metaclass $ClassName$ of the metamodel $MM_A$)
To alleviate the complexity of metamodel's structures and enable model manipulation operations without a deep understanding of the underlying concepts, we rely on Swagger\footnote{\url{https://swagger.io}}, a specification to document and present REST API endpoints.

Our middleware supports a range of model manipulation operations, as detailed in Table~\ref{tab:operations}. 
These operations are executed through either the code created by the generator model commented above or the reflection API provided by EMF.
We also employed Java reflection, thus allowing us to dynamically inspect and modify runtime attributes, including classes, interfaces, fields, and methods of generated code within the \textit{Code Repository}.

\begin{table*}[]
    \caption{Operations supported by our middleware}
    \label{tab:operations}
    \resizebox{\textwidth}{!}{%
    \begin{tabular}{@{}lllll@{}}
    \toprule
    \textbf{Id} & \textbf{Operation} & \textbf{Description}                                                                                                                                                                                                  & \textbf{\begin{tabular}[c]{@{}l@{}}HTTP \\ Method\end{tabular}} & \textbf{Endpoint}                                                                                                                                                                                                                                                                                                                                                                       \\ \midrule
    \textbf{1}  & \textit{READ}      & \begin{tabular}[c]{@{}l@{}}Read all elements from models that conform \\ to the same metamodel\end{tabular}                                                                                                           & \textit{GET}                                                    & .../\{packageName\}/\{className\}/all                                                                                                                                                                                                                                                                                                                                                   \\ \midrule
    \textbf{2}  & \textit{READ}      & \begin{tabular}[c]{@{}l@{}}Filter an element of a specific model, filtered \\ by a key attribute and its corresponding value.\end{tabular}                                                                            & \textit{GET}                                                    & \begin{tabular}[c]{@{}l@{}}.../\{packageName\}/\{className\}/\{containmentName\}/search?\\ attributeName=\textless{}param1\textgreater{}\&attributeValue=\textless{}param2\textgreater{}\end{tabular}                                                                                                                                                                                   \\ \midrule
    \textbf{3}  & \textit{ADD}       & Add an element to a specific model                                                                                                                                                                                    & \textit{POST}                                                   & .../\{packageName\}/\{parentClass\}/\{childClass\}/\{xmiFileName\}/newElement                                                                                                                                                                                                                                                                                                           \\ \midrule
    \textbf{4}  & \textit{ADD}       & \begin{tabular}[c]{@{}l@{}}Add an element to a specific class by specifying \\ the class with which it should be associated.\end{tabular}                                                                             & \textit{POST}                                                   & \begin{tabular}[c]{@{}l@{}}.../\{packageName\}/\{className\}/\{xmiFileName\}/addExisting?\\ parentContainmentName=\textless{}param1\textgreater{}\&attributeNameToMatch=\textless{}param2\textgreater{}\&\\ attributeValueToMatch=\textless{}param3\textgreater{}\&childClassName=\textless{}param4\textgreater{}\&\\ childContainmentName=\textless{}param5\textgreater{}\end{tabular} \\ \midrule
    \textbf{5}  & ADD                & \begin{tabular}[c]{@{}l@{}}In the scenario of a bidirectional relationship, \\ add an element in a specific model by setting \\ the field reference to point to an instance \\ of another class type.\end{tabular} & POST                                                            & \begin{tabular}[c]{@{}l@{}}.../\{packageName\}/\{parentClassName\}/\{childClassName\}/\{xmiFileName\}\\ /newEopposite?fieldType=\textless{}param1\textgreater{}\end{tabular}                                                                                                                                                                                                            \\ \midrule
    \textbf{6}  & \textit{UPDATE}    & \begin{tabular}[c]{@{}l@{}}Update the element based on the key attribute \\ and its corresponding value\end{tabular}                                                                                                  & \textit{PUT}                                                    & \begin{tabular}[c]{@{}l@{}}.../\{packageName\}/\{className\}/\{xmiFileName\}/update?\\ attributeName=\textless{}param1\textgreater{}\&attributeValue=\textless{}param2\textgreater{}\&updatedValue=\textless{}param3\textgreater{}\end{tabular}                                                                                                                                         \\ \midrule
    \textbf{7}  & \textit{DELETE}    & \begin{tabular}[c]{@{}l@{}}Delete the element based on the key attribute \\ and its corresponding value\end{tabular}                                                                                                  & \textit{DELETE}                                                 & \begin{tabular}[c]{@{}l@{}}.../\{packageName\}/\{className\}/\{xmiFileName\}/deleteByAttribute?\\ attributeName=\textless{}param1\textgreater{}\&attributeValue=\textless{}param2\textgreater{}\end{tabular}                                                                                                                                                                            \\ \midrule
    \textbf{8}  & \textit{DELETE}             & \begin{tabular}[c]{@{}l@{}}Delete an entire model by indicating the model\\ name to be deleted\end{tabular}                                                                                                           & \textit{DELETE}                                                          & .../\{packageName\}/\{className\}/\{xmiFileName\}/deleteClassByXMI                                                                                                                                                                                                                                                                                                                      \\ \bottomrule
    \end{tabular}
    }
\end{table*}

\section{Related Work}
\label{sec:related}

Several efforts have been made to combine MDE with Web Engineering. 
One of the prominent examples is EMF-REST~\cite{DBLP:conf/sac/Ed-DouibiIGTC16}, a model-based stand-alone generation process that creates RESTful APIs for EMF models.
However, they do not provide dynamic model management capabilities, which allow for dynamic configuration of the RESTful API as the metamodel evolves, as we target in our approach.

Another example is Sirius~\cite{viyovic2014sirius}, a framework that enables users to create domain-specific modeling editors by defining graphical representations and interactions within the Eclipse platform.
However, our approach diverges from this work in several key aspects that enhance its accessibility and flexibility. 
Firstly, we leverage widely known and existing technologies like RESTful APIs and JSON, which are familiar even to those without specialized modeling expertise. 
Additionally, our method offers hybrid solutions that extend the capability of querying across an entire repository of models, not just individual models. 

Other works include approaches that extend the utility of models and model transformations for designing and semi-automatically producing web applications, focusing on data, navigation, and presentation models~\cite{steinberg2008emf, koch2012requirements, brambilla2010tools, qafmolla2010automation, schauerhuber2006bridging}.
However, creating RESTful APIs remains significantly underdeveloped~\cite{maximilien2007domain, rivero2013mockapi, tavares2013model}. 
These methods typically require the designer to outline the API using a specialized domain-specific language (DSL), leading to only a partial API generation. 
Unlike these existing approaches, our method is designed to generate a full RESTful API directly from any data model, creating a middleware that simplifies interactions with the foundational models, bypassing the complex technical details commonly associated with utilizing MDE models.

\section{Conclusion}
\label{sec:conclusion}

In this paper, we have outlined an approach to dynamic provisioning of RESTful services for model management. 
These services can interact with any kind of model and can be used to build a modeling platform providing modeling-as-a-service.
Furthermore, the approach relies on well-known technologies, such as EMF, Spring Boot, REST APIs, and Swagger, thus ensuring its accessibility and usability.
The approach has been designed as a middleware that exposes REST APIs for distributed model management, bridging the gap between clients (e.g., web UI) and modeling infrastructure.
As a work in progress, a first implementation of our approach can be found at~\cite{repository}.

In future work, we plan to extend our approach to fully support model validation, which may imply using transactions when manipulating models.
We are also interested in exploring the distributed storage of models and how to provide transparent access to models stored in different repositories.
Finally, we plan to evaluate our approach in other domains, such as automotive, aerospace, and IoT, to assess its applicability in different contexts.

\bibliographystyle{IEEEtran}
\bibliography{icws_short}

\end{document}